\documentclass[10pt,twocolumn,superscriptaddress,showpacs,preprintnumbers,amsmath,amssymb,prl,aps]{revtex4}
\usepackage{graphicx,bm}
\usepackage{amssymb,amsmath}

\begin{document}

\title{Detection of Time-Reversal Symmetry Breaking in the Non-Centrosymmetric Superconductor Re$_{6}$Zr Using Muon-Spin Spectroscopy}

\author{R. P. Singh}
\affiliation{Physics Department, University of Warwick, Coventry, CV4 7AL, United Kingdom}

\author{A. D. Hillier}
\affiliation{ISIS facility, STFC Rutherford Appleton Laboratory, Harwell Science and Innovation Campus, Oxfordshire, OX11 0QX, United Kingdom}

\author{B. Mazidian}
\affiliation{H. H. Wills Physics Laboratory, University of Bristol, Tyndall Avenue, Bristol BS8 1TL, United Kingdom}
\affiliation{ISIS facility, STFC Rutherford Appleton Laboratory, Harwell Science and Innovation Campus, Oxfordshire, OX11 0QX, United Kingdom}

\author{J. Quintanilla}
\affiliation{ISIS facility, STFC Rutherford Appleton Laboratory, Harwell Science and Innovation Campus, Oxfordshire, OX11 0QX, United Kingdom}
\affiliation{SEPnet and Hubbard Theory Consortium, School of Physical Sciences, University of Kent, Canterbury CT2 7NH, United Kingdom}

\author{J. F. Annett}
\affiliation{H. H. Wills Physics Laboratory, University of Bristol, Tyndall Avenue, Bristol BS8 1TL, United Kingdom}

\author{D. McK. Paul}
\affiliation{Physics Department, University of Warwick, Coventry, CV4 7AL, United Kingdom}

\author{G. Balakrishnan}
\affiliation{Physics Department, University of Warwick, Coventry, CV4 7AL, United Kingdom}

\author{M. R. Lees}
\email[]{m.r.lees@warwick.ac.uk}
\affiliation{Physics Department, University of Warwick, Coventry, CV4 7AL, United Kingdom}

\begin{abstract}
We have investigated the superconducting state of the non-centrosymmetric compound Re$_{6}$Zr using magnetization, heat capacity, and muon-spin relaxation/rotation ($\mu$SR) measurements. Re$_{6}$Zr has a superconducting transition temperature, $T_c = 6.75\pm0.05$~K.  Transverse-field $\mu$SR experiments, used to probe the superfluid density, suggest an $s$-wave character for the superconducting gap. However, zero and longitudinal-field $\mu$SR data reveal the presence of spontaneous static magnetic fields below $T_c$ indicating that time-reversal symmetry is broken in the superconducting state and an unconventional pairing mechanism. An analysis of the pairing symmetries identifies the ground states compatible with time-reversal symmetry breaking.
\end{abstract}


\maketitle
\section{introduction}

The symmetry of a material plays a fundamental role in determining its physical properties. Symmetry breaking can modify the physics of a system and produce new and unusual behavior. Superconductivity is one of the best examples of a symmetry breaking phenomenon. In conventional superconductors gauge symmetry is broken, while in unconventional superconductors other symmetries may also be broken.

There has recently been a great deal of interest in non-centrosymmetric superconductors (NCS) due to the complex nature of their superconducting properties~\cite{Bauer2012}. The lack of inversion symmetry in these materials induces an anti-symmetric spin-orbit coupling (SOC)~\cite{Rashba,Rashba1} which can lift the degeneracy of the conduction band electrons and may cause the superconducting pair wavefunction to contain mixed singlet-triplet spin states. 
This mixed pairing can lead non-centrosymmetric superconductors to display significantly different properties from conventional superconducting systems e.g. nodes in the superconducting gap and upper critical fields exceeding the Pauli limiting field~\cite{Bauer2012}. In addition, some of these systems display  time-reversal symmetry (TRS) breaking.

One of the most direct ways of confirming the presence of an unconventional superconducting state is muon spectroscopy~\cite{Schenck, Lee, Yaouanc}. In a muon spectroscopy experiment, 100\% spin-polarized positive muons are implanted one at a time into a sample. After coming to rest the muon spin precesses in the local magnetic environment. The muons decay with a half-life of $2.2~\mu$s, emitting a positron preferentially in the direction of the muon spin at the time of decay. The number of positrons are recorded as a function of time in forward, $N_F(t)$, and backward, $N_B(t)$, detectors. The time evolution of the muon polarization can be obtained by examining the normalized difference of these two functions via the asymmetry function $A(t)= \frac{N_B(t)-N_F(t)}{N_B(t)+N_F(t)}$. This technique can accurately determine the magnetic penetration depth and hence the temperature dependence of the superfluid density, yielding information on the symmetry of the superconducting gap. Muon spectroscopy can also be used to unambiguously establish the onset of time-reversal symmetry breaking in superconductors. The magnetic moments associated with the Cooper pairs are non-zero in such superconductors and a local alignment of these moments produces spontaneous, but extremely small, internal magnetic fields~\cite{sigrist}. Muon spin relaxation/rotation ($\mu$SR) is especially sensitive to small changes in internal fields and can easily measure fields of $\sim10~\mu$T which correspond to moments that are just a few hundredths of a $\mu_B$.

Time-reversal symmetry breaking is rare and has only been observed directly in a few unconventional superconductors, e.g. Sr$_{2}$RuO$_{4}$~\cite{Sr2RuO4,OpticalSr2RuO4}, UPt$_{3}$ and (U,Th)Be$_{13}$~\cite{UPt3,OneUPt3,TwoUPt3,UThBe}, (Pr,La)(Os,Ru)$_{4}$Sb$_{12}$~\cite{PrOsSb,PrOsRuSb}, PrPt$_{4}$Ge$_{12}$~\cite{PrPt4Ge12}, and LaNiGa$_{2}$~\cite{LaNiGa2}. The possibility of singlet-triplet pairing in non-centrosymmetric superconductors makes them prime candidates to exhibit TRS breaking. To date, however, the only NCS reported to show TRS breaking is LaNiC$_{2}$~\cite{LaNiC2, SrPtAs}. In this material, symmetry analysis implies that the superconducting instability is of the nonunitary triplet type, with a spin-orbit coupling that is comparatively weak and with mixing of singlet and triplet pairing being forbidden by symmetry~\cite{LaNiC2Theo}. 

Several other non-centrosymmetric superconductors including Nb$_{0.18}$Re$_{0.82}$~\cite{NbRe}, Mo$_{3}$Al$_{2}$C~\cite{Mo3Al2C}, Li$_2$(Pd,Pt)$_3$B~\cite{Yuan2006, Nishiyama2007, Takeya2007, Harada2012}, Ca(Ir,Pt)Si$_{3}$~\cite{CaIr}, LaRhSi$_{3}$~\cite{LaRhSi3}, Mg$_{10}$Ir$_{19}$B$_{16}$~\cite{Mg10Ir19B16}, and Re$_{3}$W~\cite{ReW} have been studied by magnetization, transport, and heat capacity measurements and some have been shown to exhibit unconventional superconducting behavior including triplet pairing~\cite{Nishiyama2007} and upper critical fields close to the Pauli limit~\cite{Mo3Al2C}.

\begin{figure}[b]
\centering
\includegraphics[width=.8\columnwidth]{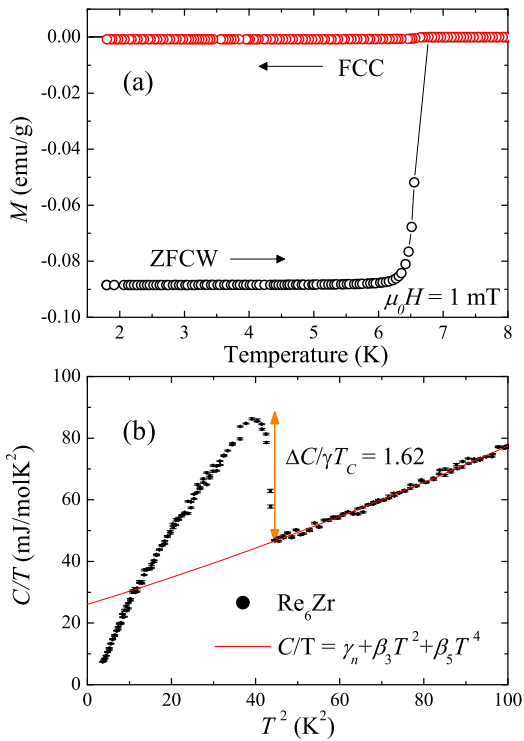}
\caption{(Color online) (a) Temperature dependence of the magnetization collected in zero-field-cooled warming (ZFCW) and field-cooled cooling (FCC) modes in an applied field $\mu_0H=1.0$~mT. (b) $C/T$ versus $T^{2} $ for Re$_{6}$Zr. The line is a fit to the data using $C/T=\gamma_n+\beta_3T^{2}+\beta_5T^{4}$ where $\gamma_n T$ and $\beta_3T^3+\beta_5T^5$ are the normal state electronic and lattice contributions to the specific heat respectively. $\gamma_n=26\pm2$~mJmol$^{-1}$K$^{-2}$ and the Debye temperature calculated from $\beta_3$ is $319\pm9$~K.}
\label{HCdata}
\end{figure}

Studies using $\mu$SR have also been performed on the last four of these compounds~\cite{LaRhSi3,muon-MgIrB,muon-ReW, muon-CaIrPtSi}. However, no spontaneous fields were observed in the superconducting state of any of these materials indicating that the breaking of TRS is either absent or undetectable in these materials. The failure to detect TRS breaking in a number of NCS containing heavy transition metals in which SOC is normally expected to be strong and in which mixed spin-singlet spin-triplet pairing may be allowed, raises the possibility that TRS breaking may be absent in this particular class of superconductors. In order to address this question we have begun a systematic investigation of the superconducting properties of a number of non-centrosymmetric materials with the body-centered cubic $\alpha$-Mn structure~\cite{Matthias, Gladyshevskii}, containing the heavy transition-metal Re. In this Letter we report the results of $\mu$SR measurements on one of these materials, Re$_{6}$Zr~\cite{Savitskiei, Matthias, Gladyshevskii}. Zero and longitudinal-field $\mu$SR reveal that spontaneous magnetic fields develop at the superconducting transition temperature, confirming the presence of TRS breaking in the superconducting state of Re$_{6}$Zr. Transverse-field $\mu$SR data and a theoretical analysis of the possible pairing states in Re$_6$Zr demonstrate that there is a mixing of spin-singlet and spin-triplet pairing in this non-centrosymmetric superconducting compound.

\begin{figure}[tb]
\centering
\includegraphics[width=.8\columnwidth]{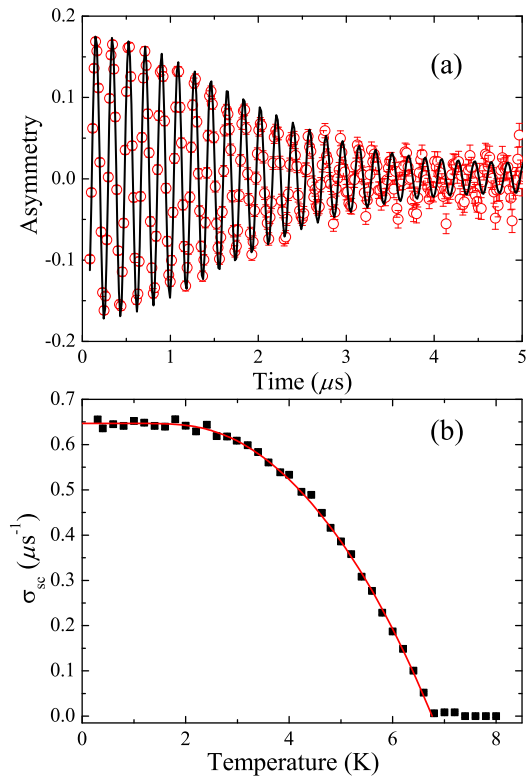}
\caption{(Color online) (a) Transverse-field muon time spectra for Re$_{6}$Zr at $T = 0.3$~K in a magnetic field 40~mT. The solid line shows a fit to the data using Eq.~\ref{equ1}. For clarity, only the data from one of the two virtual detectors is shown (b) Temperature dependence of the superconducting muon spin depolarization rate $\sigma_{sc}$ collected in an applied magnetic field $\mu_0H = 40$~mT. The solid line shows a fit to an isotropic $s$-wave gap.}
\label{TFdata}
\end{figure}

Polycrystalline samples of Re$_{6}$Zr were prepared by arc melting stoichiometric quantities of high purity ($5N$) Zr and Re in a tetra-arc furnace (Cyberstar, Grenoble, France) under an argon ($5N$) atmosphere on a water-cooled copper hearth. The sample buttons were melted and flipped several times to ensure phase homogeneity. The observed weight loss during the melting was negligible. Powder x-ray diffraction (PXRD) data confirmed the sample have the $\alpha$-Mn crystal structure (space group $I\bar{4}3m$, No. 217). PXRD and energy dispersive x-ray spectroscopy showed that no other phases, including oxides and carbides, are present in the sample to within the limits of these analysis techniques.

The materials were characterized using magnetization, $M$, and heat capacity, $C$, measurements. These measurements indicate that Re$_{6}$Zr is a bulk superconductor with a superconducting transition temperature, $T_c$, of $6.75\pm0.05$~K (Fig.~\ref{HCdata}(a)). This is less than the 7.4~K reported previously~\cite{Matthias}. $\Delta C/\gamma_n T_c=1.62\pm0.09$ is larger than the $1.43$ expected for a conventional BCS superconductor (Fig.~\ref{HCdata}b). This suggests the presence of an enhanced electron-phonon coupling compared with a BCS superconductor~\cite{Re6Zr}. Using the Werthamer-Helfand-Hohenberg expression~\cite{WHH}, $H_{c2}(0)=0.693T_c(dH_{c2}/dT)_{T_c}$ we estimate $\mu_0H_{c2}(0)=12.20\pm0.06$~T which is close to the value of the Pauli paramagnetic limiting field, $\mu_0H_{\mathrm{Pauli}} = 1.84T_c$ (in tesla) of $12.35\pm0.09$~T. 

Zero-field, longitudinal-field, and transverse field $\mu$SR experiments were carried out using the MuSR spectrometer at the ISIS pulsed muon facility. A detailed description of the different instrumental geometries can be found in Ref.~\cite{Lee}. A powdered sample of Re$_{6}$Zr was mounted on a sample plate made of 99.995\% silver. The sample holder and sample were placed in a Helium-3 cryostat with a temperature range of 0.3-10~K. The stray fields at the sample position were canceled to a level of $1\mu$T by an active compensation system. 

Transverse field $\mu$SR (TF-$\mu$SR) experiments were performed in the superconducting mixed state in applied fields between 40 and 60 mT, well above the $\mu_0H_{c1}(0)=8\pm1$~mT of this material~\cite{Re6Zr}. In this geometry, the 64 detectors in the MuSR spectrometer are combined in software to provide two orthogonal virtual detectors, each with a phase offset $\phi$. Data were collected in field-cooled (FC) mode where the magnetic field was applied above the superconducting transition and the sample was then cooled to base temperature. 

\begin{figure}[tb]
\centering
\includegraphics[width=0.8\columnwidth]{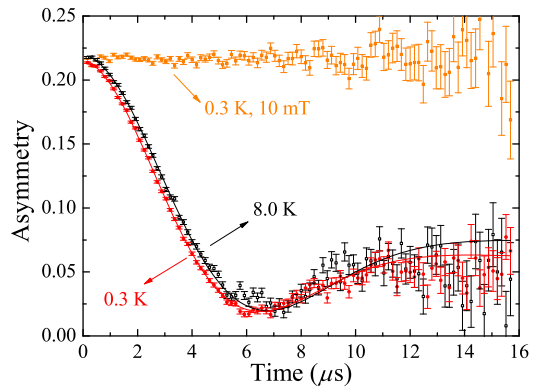}
\caption{(Color online) Zero-field $\mu$SR time spectra for Re$_{6}$Zr collected at 0.3 (closed symbols) and 8.0~K (open symbols) are shown together with lines that are least squares fits to the data using Eq.~\ref{KT2}. These spectra taken below and above $T_c$ are representative of the data collected. A LF-$\mu$SR time spectrum taken in an applied field of 10~mT at 0.3~K is also shown.}
\label{LFdata}
\end{figure}

Figure~\ref{TFdata}(a) shows a typical TF-$\mu$SR precession signal for Re$_{6}$Zr with an applied field of 40~mT at 300~mK. The signal decays with time because of the inhomogeneous field distribution of the flux-line lattice. The TF spectra were fitted using a sinusoidal oscillating function with a Gaussian relaxation, and an oscillatory background term arising from the muons implanted directly into the silver sample holder that do not depolarize.
\begin{eqnarray}
\label{equ1}
G\left(t\right)=A_{1}\exp\left(-\frac{\sigma_{sc}^{2}t^{2}}{2}\right)\cos\left(2\pi v_{1}t+\phi_1\right) \nonumber\\ 
+A_{2}\cos\left(2\pi v_{2}t+\phi_2\right).
\end{eqnarray}
$\sigma_{sc}$ exhibits no field dependence between 40 and 60 mT, as expected for a material with a $\mu_0H_{c2}(0)$ of 12.2~T. Figure~\ref{TFdata}b shows the $T$ dependence of the muon depolarization rate which can be directly related to the superfluid density~\cite{Lee, Uemura, Sonier}. From this, the nature of the superconducting gap can be determined. The data can be well modeled by a single isotropic $s$-wave gap of $1.21\pm0.08$~meV. This gives a gap of $2\Delta/k_{B}T_c = 4.2\pm0.3$, which is higher than the 3.53 expected for BCS  superconductors. This is a further indication of the enhanced electron-phonon coupling in the superconducting state.

\begin{figure}[tb]
	\centering
\includegraphics[width=0.8\columnwidth]{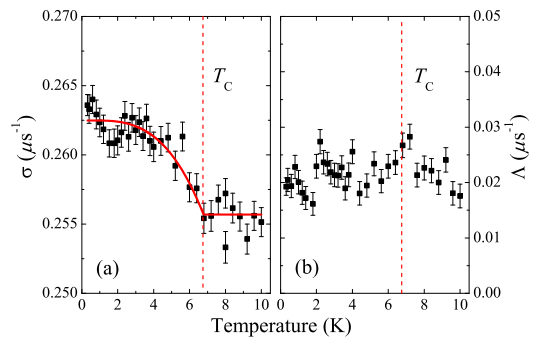}
\caption{(Color online) (a) Temperature dependence of $\sigma$ for Re$_{6}$Zr in zero field which clearly shows the spontaneous fields appearing at $T_c=6.78$~K. The solid line is a fit to the data (reduced $\chi_\nu^2=1.02$) using an approximation~\cite{Carrington} to the BCS order parameter for $\sigma$. (b) The electronic relaxation rate, $\Lambda$, versus temperature in zero field shows no temperature dependence.}
\label{LFresults}
\end{figure}

Zero-field (ZF) relaxation data were collected at several temperatures between 0.3 and 10~K. Figure~\ref{LFdata} shows the spectra collected at 8 and 0.3~K, i.e. above and below $T_c$ respectively. In these relaxation experiments, any muons stopped in the silver sample holder give a time independent background. There is no indication of a precessional signal over the entire temperature range studied, ruling out the possibility of a large internal field and hence long-range magnetic order in Re$_{6}$Zr. In the absence of atomic moments, the only possibility is that the muon spin relaxation is due to static, randomly oriented local fields associated with the nuclear moments at the muon site. This depolarization can be described by the Kubo-Toyabe function 
\begin{equation}
\label{KT1}
G^{\rm{KT}}\left(t\right)=\left(\frac{1}{3}+\frac{2}{3}\left(1-\sigma^{2}t^{2}\right)\exp\left(-\frac{\sigma^{2}t^{2}}{2}\right)\right)
\end{equation} and our spectra for Re$_{6}$Zr are well described by the function 
\begin{equation}
\label{KT2}
G\left(t\right)=A_{0}G^{\rm{KT}}\left(t\right)\exp\left(-\Lambda t\right)+A_{\rm{bckgrd}},
\end{equation}
where $A_{0}$ is the initial asymmetry, $A_{\rm{bckgrd}}$ is the background, and $\Lambda$ is the electronic relaxation rate.

The parameters $A_{0}$, $\Lambda$, and $A_{\rm{bckgrd}}$ are found to be temperature independent. Only the depolarization rate, $\sigma$, shows any temperature dependence increasing with decreasing temperature below $T_c$ (see Fig.~\ref{LFresults}). Such behavior is unusual. To date, such a change in $\sigma$ has only been observed in superconducting PrOs$_{4}$Sb$_{12}$ and PrPt$_{4}$Ge$_{12}$~\cite{PrOsSb, PrPt4Ge12}, and more recently in LaNiGa$_{2}$~\cite{LaNiGa2}, none of which are NCS. The temperature dependence of $\sigma$ agrees with a BCS order parameter. This increase in $\sigma$ can be explained in terms of a signature of a coherent internal field with a very low frequency as discussed by Aoki~\textit{et al}.~\cite{PrOsSb} for PrOs$_{4}$Sb$_{12}$. They suggest that the field distribution is Gaussian in nature compared to Sr$_{2}$RuO$_{4}$, where $\Lambda$ exhibits a spontaneous increase below $T_c$ due to a Lorentzian-type field distribution. In order to rule out the possibility that the increased relaxation in the superconducting state is due to extrinsic effects such as impurities, measurements were performed in weak longitudinal fields (Fig.~\ref{LFdata}). A longitudinal magnetic field of just 10~mT removes any relaxation due to the spontaneous fields and is sufficient to fully decouple the muons from this relaxation channel. This in turn shows that the associated magnetic fields are in fact static or quasi-static on the time scale of the muon precession. These observations confirm time-reversal symmetry breaking in Re$_{6}$Zr in the superconducting state.

In noncentrosymmetric superconductors, spin-orbit coupling can lead to mixed singlet-triplet pairing~\cite{Rashba1},
\begin{equation}
\hat{\Delta}({\bf k}) = i\left[ \Delta(\mathbf{k}) + \mathbf{d}(\mathbf{k}) \cdot \hat{\boldsymbol{\sigma}} \right] \hat{\sigma}_y.
\label{mixed_singlet_triplet}
\end{equation}
This in itself, however, does not suffice to yield broken time-reversal symmetry. Take for instance the case of Li$_2$Pd$_{3-x}$Pt$_x$B ($0 \leq x \leq 3$). In this system, SOC induces a triplet component, $i\mathbf{d}(\mathbf{k}) \cdot \hat{\boldsymbol{\sigma}} \hat{\sigma}_y$, that can be as small as $\sim$25\% or as large as $\sim$167\% of the total gap energy for $x=0$ and $3$, respectively~\cite{Yuan2006, Nishiyama2007, Takeya2007, Harada2012}. The pairing mechanism in this family is believed to be conventional for all values of $x$ and pairing is thus thought to preserve all the lattice symmetries including TRS. In contrast, the breaking of time-reversal symmetry necessitates a superconducting instability in a higher-dimensional irreducible representation of the crystal's point group. Such a state breaks additional lattice symmetries in addition to TRS. This rules out any conventional pairing mechanism, as it would necessarily lead to an order parameter in the most symmetric irreducible representation of the crystal point group, $A_1$. 

The results presented here make Re$_6$Zr only the second NCS where broken TRS has been observed directly. Moreover the other known example, LaNiC$_2$~\cite{LaNiC2}, has a crystal point group, $C_{2v}$, whose irreducible representations are all one-dimensional. It has thus been argued that for LaNiC$_2$ the orbital and spin degrees of freedom must be approximately decoupled at the instability, so that pairing takes place in an almost purely triplet channel, rather than a mixed singlet-triplet one~\cite{LaNiC2Theo}. This scenario has been supported by the observation of broken TRS in centrosymmetric LaNiGa$_2$~\cite{LaNiGa2}. 

In comparison to the elements in LaNiC$_2$, Re has a much higher atomic number and we expect Re$_6$Zr to feature strong singlet-triplet mixing. The relevant point group, $T_d$, features one two-dimensional irreducible representation ($E$) and two three-dimensional irreducible representations ($F_1$ and $F_2$)~\cite{LandauLifshitz}. Minimization of generic Landau free energies for superconducting instabilities in the $E$  channel~\cite{Annett1990} reveals three possible ground states for this two-dimensional irreducible representation. 
 
Of those, only one ground state breaks time-reversal symmetry. Its gap matrix has the form of Eq.~(\ref{mixed_singlet_triplet}) with the scalar $\Delta(\mathbf{k})$ and vector $\mathbf{d}(\mathbf{k})$ given, respectively, by
\begin{eqnarray}
\Delta(\mathbf{k}) & = & 2Z^2-X^2-Y^2+i\left(X^2-Y^2\right), \\
\mathbf{d}(\mathbf{k}) & = & \left[X(Y^2-Z^2),Y(X^2-Z^2),Z(X^2-Y^2)\right] \nonumber \\ &&  \times \left[2Z^2-X^2-Y^2+i\left(X^2-Y^2\right)\right].
\end{eqnarray}
The $F_1$ and $F_2$ channels each have more ground states that break TRS, of which we give here two examples: 
\begin{eqnarray}
\Delta(\mathbf{k}) & = & (YZ+iXZ)\\
&&\times (X^2-Y^2)(Y^2-Z^2)(Z^2-X^2), \nonumber \\
\mathbf{d}(\mathbf{k}) & = & \left(1,i,0\right) XYZ
\end{eqnarray}
for the $F_1$ irreducible representation; and 
\begin{eqnarray}
\Delta(\mathbf{k}) & = &  \left(Y+iX\right)Z,\\
\mathbf{d}(\mathbf{k}) & = & \left(1,i,0\right)XYZ \nonumber \\
&& \times (X^2-Y^2)(Y^2-Z^2)(Z^2-X^2)
\end{eqnarray}
for $F_2$. Here, as usual, $X=X(\mathbf{k})$, $Y=Y(\mathbf{k})$ and $Z=Z(\mathbf{k})$ denote three real functions that transform in the same way as the three components of the wave vector $\mathbf{k}$ under the symmetry operations of the point group. 

In summary, we have have studied the non-centrosymmetric superconducting compound Re$_{6}$Zr using $\mu$SR. TF data reveal that the temperature dependence of the muon depolarization rate can be described using an $s$-wave model. The presence of spontaneous fields below $T_c$, however, provide convincing evidence for time-reversal symmetry breaking in this material and an unconventional pairing mechanism. As a result of the presence of Re, strong singlet-triplet mixing is expected in this material. Ground states consistent with the crystallographic symmetry and the observation of TRS breaking are discussed.

\begin{acknowledgments}
The authors would like to thank T. E. Orton for valuable technical support. This work is funded by the EPSRC, United Kingdom, through grant EP/I007210/1 and by HEFCE through the South-East Physics network (SEPnet). Some of the equipment used in this research was obtained through the Science City Advanced Materials project: Creating and Characterizing Next Generation Advanced Materials project, with support from Advantage West Midlands (AWM) and part funded by the European Regional Development Fund (ERDF). 
\end{acknowledgments}


\begin{thebibliography}{References}
\bibitem{SrPtAs}  SrPtAs exhibits TRS breaking~\cite{SrPtAs-ref}. This material has a centrosymmetric structure that is locally non-centrosymmetric.
\end{thebibliography}
\end{document}